# Application of Ontologies in Cloud Computing: The State-Of-The-Art

Fahim T. Imam
School of Computing
Queen's University at Kingston
Kingston, Ontario, Canada
+1 (613) 331-4657
imam@cs.queensu.ca

## ABSTRACT

This paper presents a systematic survey on existing literatures and seminal works relevant to the application of ontologies in different aspects of Cloud computing. Our hypothesis is that ontologies along with their reasoning capabilities can have significant impact on improving various aspects of the Cloud computing phenomena. Ontologies can promote intelligent decision support mechanisms for various Cloud based services. They can also provide effective interoperability among the Cloud based systems and resources. This survey can promote a comprehensive understanding on the roles and significance of ontologies within the overall domain of Cloud Computing. Also, this project can potentially form the basis of new research area and possibilities for both ontology and Cloud computing communities.

## 1. INTRODUCTION

The philosophical term 'ontology' was first adapted to Computer science by Tom Grubar [1] as an "explicit specification of conceptualization" for the Artificial Intelligence community. The term has been used and defined in various different ways by the knowledge representation and reasoning communities, ever since. An ontology represents the concepts within a domain and specifies how the concepts are related with each other through logical axioms expressed in a formal language (e.g., OWL Description Logic language by W3C for Semantic Web [12]). Typical layers of an ontological knowledge model may include conceptual layer, formal logical layer, and application layer as depicted in Figure 1.

One of the most powerful features of ontology is that it provides a way to express *explicit* knowledge of a conceptual domain from which the *implicit* new knowledge can be *inferred* through logical reasoners or inference engines [3]. There are various open source reasoners and inference engines available, which can provide automated classification and consistency checking for the concepts specified within an ontology. Some of the widely used ontological reasoners are Racer, Pallet, and Fact++ [4] etc. The idea of ontology and its reasoning capabilities have been exploited in numerous different ways in different fields of applied computer science beyond the AI community, such as in computational biology and bioinformatics, computational neuroscience and neuroinformatics, and also in now ubiquitous Semantic Web frameworks and technologies promoted by the W3C consortium. As for the Cloud Computing paradigm, the applications of ontologies are relatively new and limited, and, therefore, their full potentials are yet to be realized.

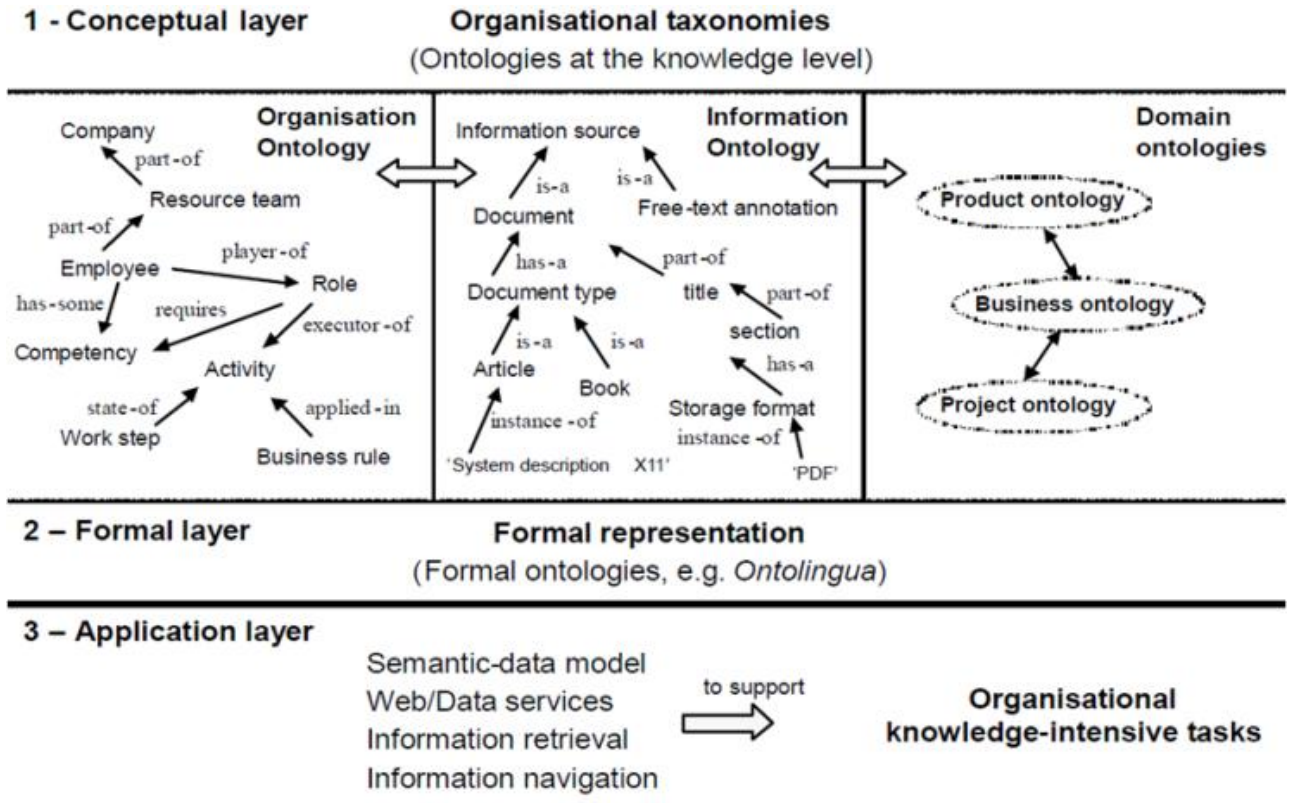

**Figure 1. Typical layers in ontological knowledge model [2].**

In this paper, we present a survey on a number of literatures and efforts published in recent years relevant to the application of ontologies within the Cloud environments. This survey will enable us to explore the overall landscape of ontologies and their applications within the broad domain of Cloud computing. The rest of the paper is organized as follows. Section 2 provides a brief discussion on the motivation and the related works. Section 3 provides the method of this survey project which includes the hypothesis, research procedure, research questions, and the classification of relevant efforts to be surveyed. Section 4 will be focused on the actual results of the surveyed literatures. The discussion in Section 5 serves as the synthesis of the overall results observed in the surveyed literatures. Finally, we will have the concluding remarks for this paper in Section 6.

## 2. MOTIVATION AND RELATED WORKS

Observing the literature corpus through the Google Scholar search reveals the fact that this kind of survey on the application of ontologies within the domain of Cloud computing is severely scarce. However, the number of recent efforts on the application of ontologies in various aspects of Cloud Computing is quite noticeable. To our best knowledge, the only related work that we could find that could qualify as a "survey" is the paper by Androcec et al. (2012) in [7]. However, the effort in [7] seems to be quite limited in terms of its scope, length (only 5 pages) and contents in order to provide a fair understanding about how ontologies are actually applied in different aspects of Cloud computing. These latter facts urged us to work on a comprehensive survey of the current state-of-the-art techniques, trends, and practices that exists within the literature corpus of ontologies and Cloud computing. The goal of this paper is to explore the key landscapes of ontologies within the domain of Cloud computing. It is hoped that this survey can potentially promote new research areas and possibilities for both ontology and Cloud Computing communities.

# 3. METHODS

## 3.1 Hypothesis

The hypothesis of this project is that ontologies along with their reasoning capabilities can have a significant impact on improving various aspects of Cloud Computing phenomena. Ontologies can promote intelligent decision support mechanisms for various cloud based services. They can also promote effective interoperability among the Cloud based systems and resources.

## 3.2 Research Procedure

Our research on this survey is based on systematic review procedures as described by Kitchenham (2004) in [10]. We have also borrowed some instructions from Biolchini et al. (2005) [11] as needed. Both articles provided a set of guidelines to address specific issues in Software Engineering researches along with a formalized process to assess and interpret all available research relevant to specific research questions, topic area, or phenomenon of interest [10, 11]. Kitchenham's *Systematic Review Procedure* provided a practical understanding of the phases of planning, conducting and reporting an effective review specific to Computer science.

## 3.3 Research Questions

For the purpose of performing a systematic survey, this paper attempts to address the following research questions in general:

- Why do we need ontologies for the Cloud? How significant the efforts are so far?
- What are the key limiting factors of the ontology-based tools and systems developed for the Cloud?
- What are the specific aspects of Cloud Computing that can be improved or simplified using ontologies and their reasoning capabilities?

All these questions will be addressed in a detailed discursive manner in Section 5.

## 3.4 Classification of Efforts

In order to provide a systematic survey, we must provide a common classification scheme for the literatures for a broad phenomena like Cloud computing. Based on the search results obtained through Google Scholar, we considered the following classification of the common themes:

1. Interoperability among the Cloud Systems
2. Symantec Models in Cloud Computing
3. Cloud Computing Security
4. Cloud Resource Management and Service Discovery

This classification is based on the specific Cloud computing issues that were targeted to improve or simplify through the application of ontologies.

# 4. RESULTS

As the title of this paper suggests, this project is expected to result in providing the current state-of-the-art understandings of ontologies and their applications in overall Cloud computing phenomena. In this section, we provide the results in terms of the issues classified in Section 3.4.

## 4.1 Cloud Interoperability

One of the biggest motivational aspects of having ontologies within the domain of Cloud computing would be to see the 'big picture' view of a universal, interoperable framework. The issue of portability and interoperability among the Cloud service providers can be seen as a critical challenge. For example, building applications targeted for Amazon's EC2 [15] can be significantly different from Google's AppEngine [16]. Similarly, data storing mechanism on Amazon's S3 [17] can be different from other cloud-based data storage facilities. Note that Amazon and Google are not the only players in the game, and the numbers of different Cloud based companies are increasing rapidly.

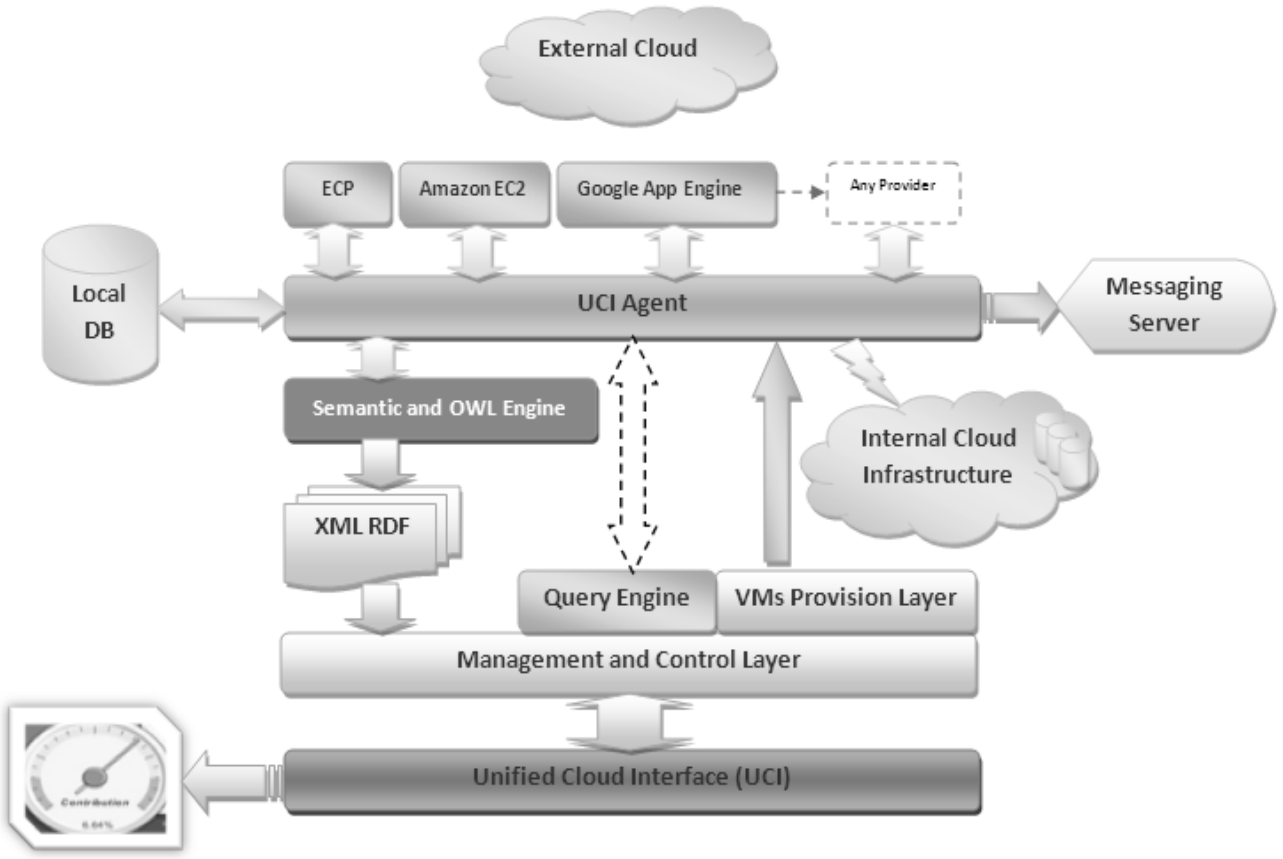

**Figure 2. Universal Cloud Interface (UFC) Architecture [5].**

The task of porting specific Cloud based applications or the data from one Cloud provider to another can become very challenging and costly, depending on the cloud platforms. Also, developing applications for a specific cloud platform cannot guarantee their portability and interoperability with other cloud platforms. Therefore, having standardized APIs can significantly reduce the efforts and costs as the developers would not need to learn any new model and APIs each time they need to port their applications and data to a new platform. Also, application developers and service providers would more likely want to support multiple different platforms to reach out to the maximum number of clients, regardless of their Cloud platforms. The initiative like Universal Cloud Interface (UCI) [5] can be seen as a hope to resolve the portability and interoperability issues, which employs the core idea of ontologies and semantic web technologies. As the UCI requirement document [6] states, "The concept is to provide a single interface that can be used to retrieve a unified representation of all multi-cloud resources and to control these resources as needed." Figure 2 provides a high-level architecture as depicted by the UCI project [5, 6]. Note that, the core of the architecture is the use of ontologies and Semantic Web technologies such as OWL [3, 12] inference engine and RDF/XML.

One of the seminal works that was conducted in order to resolve the very issue of Cloud interoperability was presented in [9] by Youseff et al. (2008). The authors provided an excellent guideline to develop a unified ontology for the overall cloud computing phenomena. The goal of the paper was to provide a unified understanding of the overall landscape of Cloud computing. The paper provides a systematic dissection of the Cloud computing field and proposes five distinct layers.

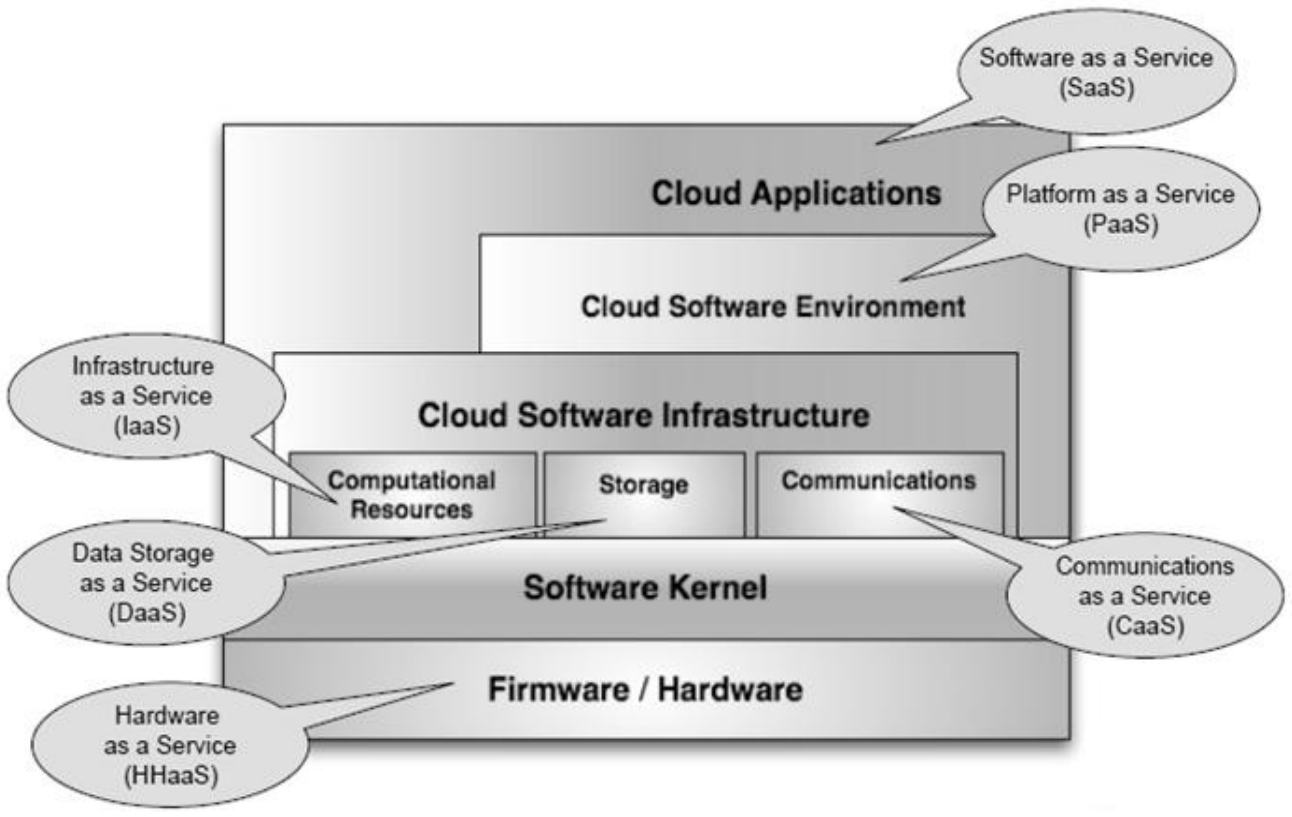

**Figure 3. The five layers of Cloud Computing Ontology proposed by Youseff et al. [9].**

As depicted in Figure 3, the five layers proposed in the paper are Cloud Application Layer, Cloud Software Environment Layer, Cloud Software Infrastructure Layer, Software Kernel, and Hardware and Firmware, with three constituent parts for the Cloud Infrastructure Layer such as Computational Resources, Storage, and Communications. The paper provides a detailed description about the inter-relation between the layers and the inter-dependencies on the preceding Cloud computing technologies. The classification of the stacks (Figure 3) were suggested based on the following rules: a. one cloud layer is higher in the stack *if* its services can be composed of the underlying layers; b. Cloud services should belong to a same layer *if* the services have the equivalent levels of abstraction.

This paper [9] has been cited over 400 times and most of the papers related to ontology and Cloud computing references this work, in one way or another. Although the paper serves as a document for comprehending the human understanding of the cloud computing phenomena, the formal aspects of ontologies were not considered explicitly. However, the layers described in the paper can be considered as the high-level concepts (or classes) in a formal cloud computing ontology. The inter-dependencies and the inter-relations specified by the paper can be encoded in a formal ontology as a set of relational object properties among different concepts within different cloud computing ontologies.

One significant contribution that exploits the formal aspects of ontologies within the domain of Cloud computing is the EU effort called the mOSAIC project (http://www.mosaic-cloud.eu). A critical goal of the mOSAIC project was to develop a common cloud ontology that could be compliant with several standards (e.g., OCCI, DMTF, and NIST etc.), platforms and services. The mOSAIC platform provides a collaborative tool for ontology development and cloud service annotations. The mOSAIC ontology provides a comprehensive collection of concepts that provides a unified formal description of different Cloud computing components, Interfaces, APIs, Requirements, SLAs, Cloud service compositions and so forth. The mOSAIC project also includes a semantic-based query execution engine (Figure 7) to support brokering, resource discovery and semantic match-making between the clients and the Cloud service providers.

The work done by Moscato et al.[18] provides a detailed analysis of mOSAIC ontology for resource annotations in Cloud environment. The ontology-based framework proposed by mOSAIC can potentially improve interoperability among existing Cloud systems, platforms and services, both from the perspectives of end-user and developers.

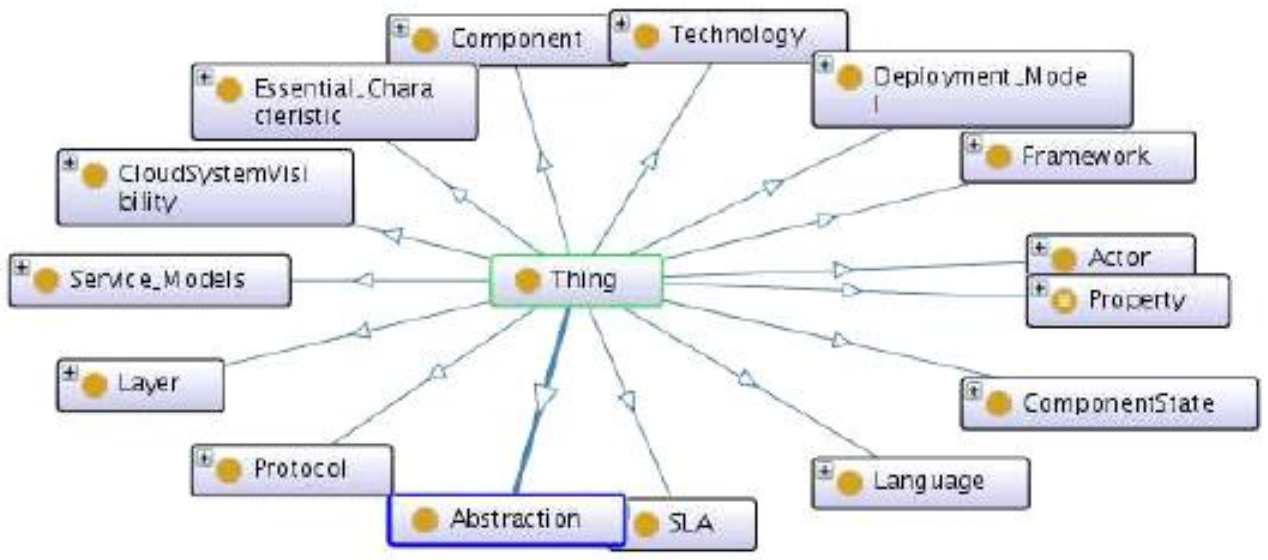

**Figure 4. The top level concepts in mOSAIC Cloud Computing Ontology [18].**

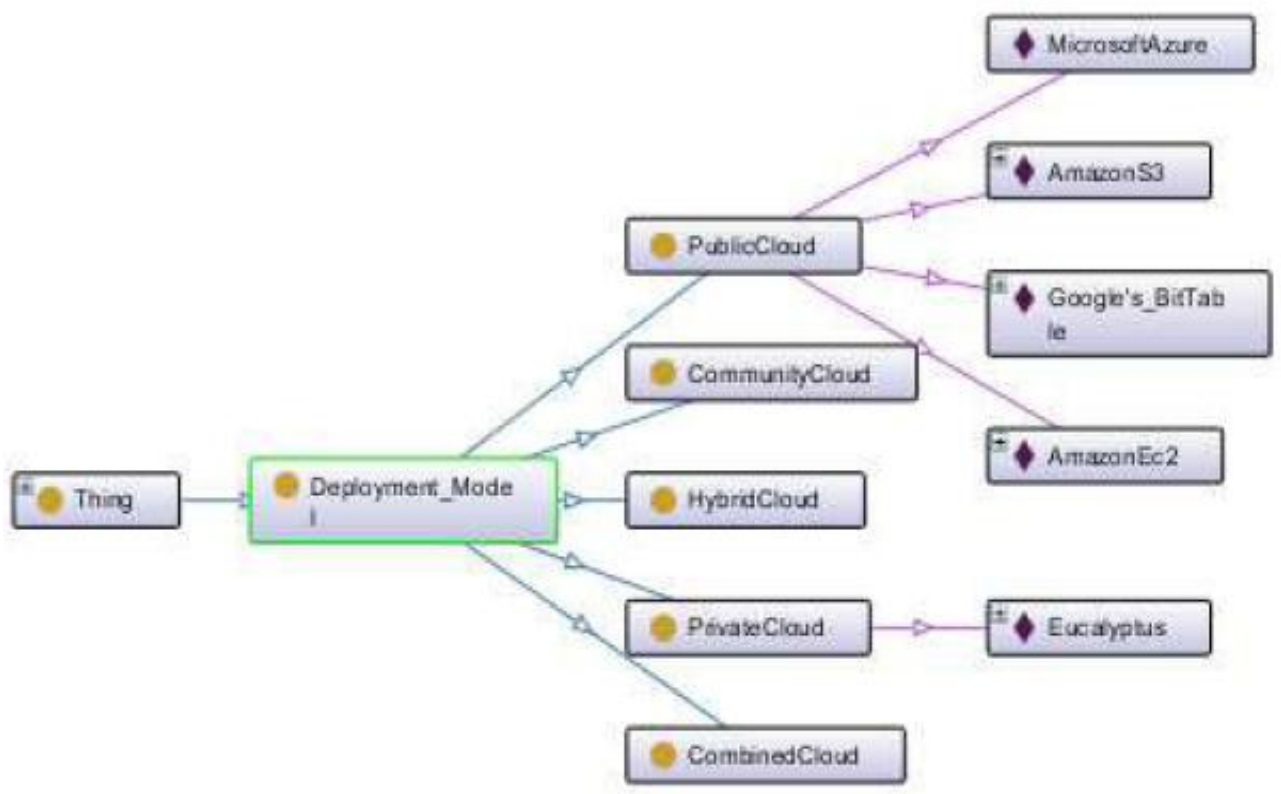

**Figure 5. Typical instantiations (marked with purple diamonds) of the public and private cloud deployment models in mOSAIC Cloud Ontology [18].**

The mOSAIC ontology is developed using W3C Standard OWL (Web Ontology Language) description logic language [12]. The top level concepts of the mOSAIC Cloud ontology are shown in Figure 4. Figure 5 depicts the typical instances of the concepts 'Public Cloud' and 'Private Cloud' in mOSAIC ontology. Table 1 contains a typical set of object properties for individuals (instances) in mOSAIC ontology.

**Table 1: Typical Object Properties for individuals used in mOSAIC ontology [18].**

| Domain | ObjectProperty | Range |
|---|---|---|
| mOSAICDataStore | developedWithLanguage | JAVA |
| mOSAICDataStore | fulfils | HighReplication |
| mOSAICDataStore | fulfils | HighConsistency |
| mOSAICDataStore | hasServiceProvidedby | Google |
| mOSAICDataStore | isOfferedByProvider | Google |
| mOSAICDataStore | isOwnedBy | Google |
| Google | guarantee | HighReplication |
| Google | guarantee | HighConsistency |
| Google | guarantee | LowConsistency |
| Google | offersAPI | GoogleAppAPI |
| Google | own | mOSAICDataStore |
| GoogleAppAPI | developedwithLanguage | JAVA |
| GoogleAppAPI | developedwithLanguiage | Python |
| GoogleAppAPI | developedwithLanguiage | Go |

Based on Table 1, Figure 6 depicts the individuals and their relationships to mOSAICDataStore. The edges with different colors in Figure 6 represent different properties from Table 1.

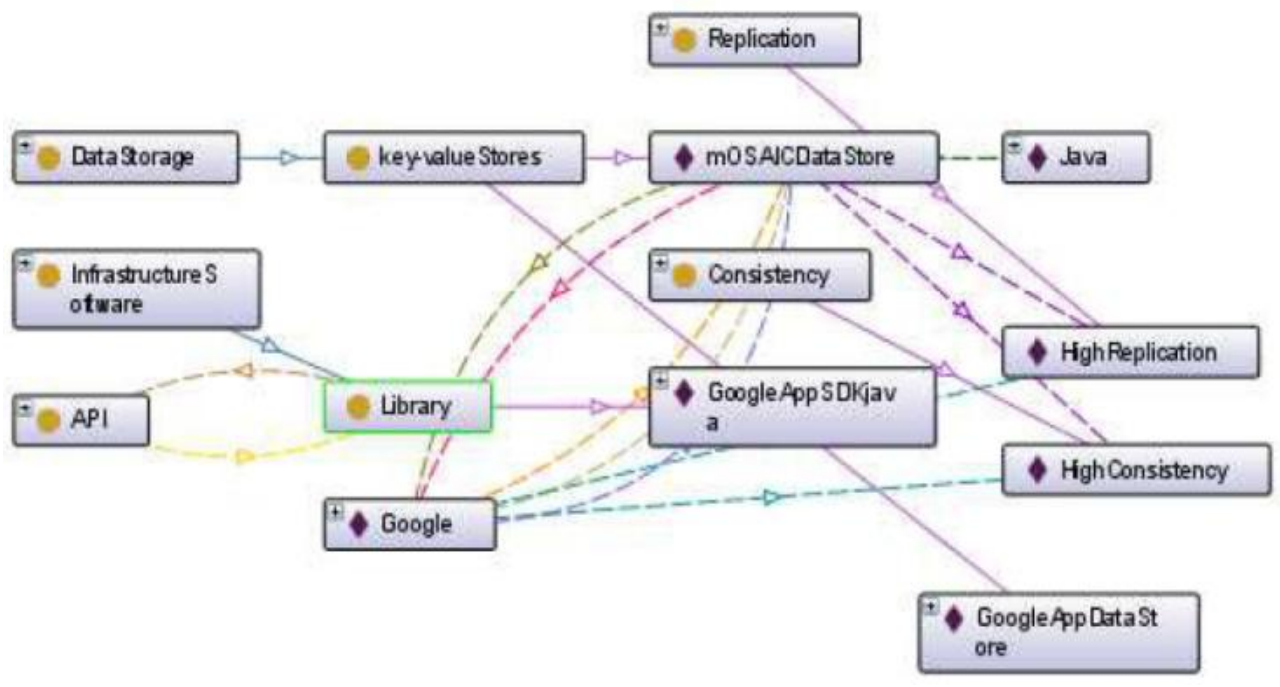

**Figure 6. The mOSAICDataStore main individuals and their relationships [18].**

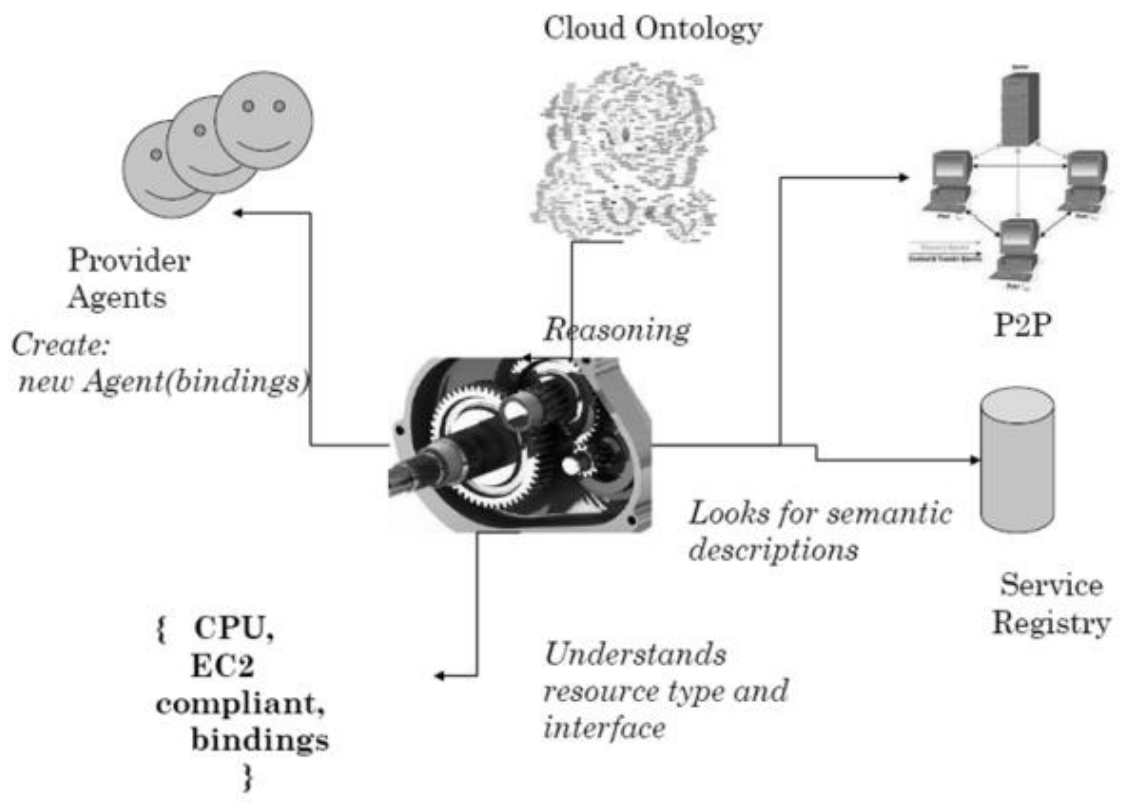

**Figure 7. The mOSAIC Cloud Semantic Query Engine.**

## 4.2 Semantic Models in Cloud Computing

One of the critical aspects of ontologies is that they can provide semantic models that can define an application domain in a precise logical manner. The idea of semantic modeling has long been associated as the core basis of knowledge representation and reasoning.

Sheth et al. (2010) explores three aspects of cloud computing where semantic models can potentially help [19]. First, semantic models can provide platform-agnostic specifications to define application functionality and quality of service (QoS) details for the cloud systems. In other words they can help to specify the functional and non-functional (e.g., SLA) components of different cloud environments. Second, semantic models can help to model the massive amount of data in the cloud to overcome the difficulty of porting data horizontally across different clouds. This can help, for example, to move or replicate data from a schema-less data store (e.g., Google Bigtable) to a schema-based data store (e.g., relational databases). The third aspect of Cloud computing where semantic models can help is the fact that they can be used to enhance the Cloud service descriptions. The authors pointed out that while exposing their operations via Web services, different Cloud systems (vendors) usually deploy different service interfaces (APIs) according to their needs. However, the operation semantics of those Web services from different vendors may represent similar operations.

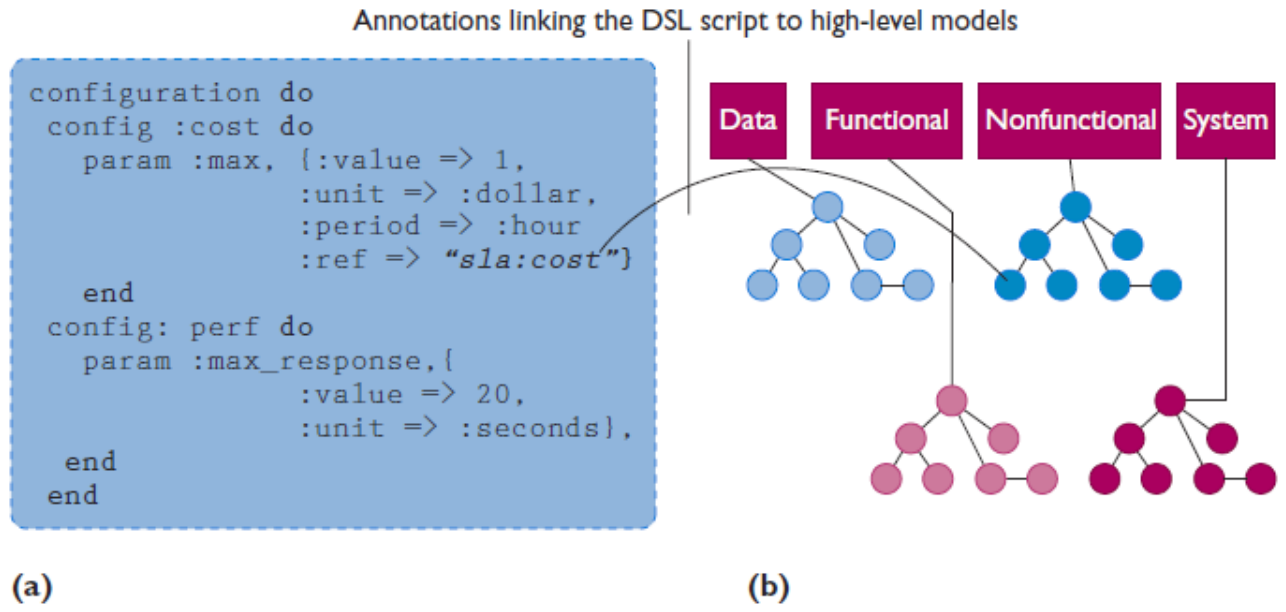

**Figure 8. Annotating (a) a domain-specific language (DSL) script with (b) high-level semantic models as depicted by Sheth et al. (2010)[19].**

In order to consolidate these differences in APIs, the authors suggested that semantic models can help to annotate the APIs with their corresponding generic operational models. This can essentially enable much desired interoperability among heterogeneous Cloud environments.

Figure 8 depicts a simple scenario where a script written in a domain specific language (DSL) is being annotated and mapped with a high-level semantic model as extracted from [19]. The annotations can link the relevant components between different levels as well. In this way, any 'programmer-driven' DSL script can accomplish ontological richness, and effectively provide a way to facilitate high-level operations while maintaining a simpler representation.

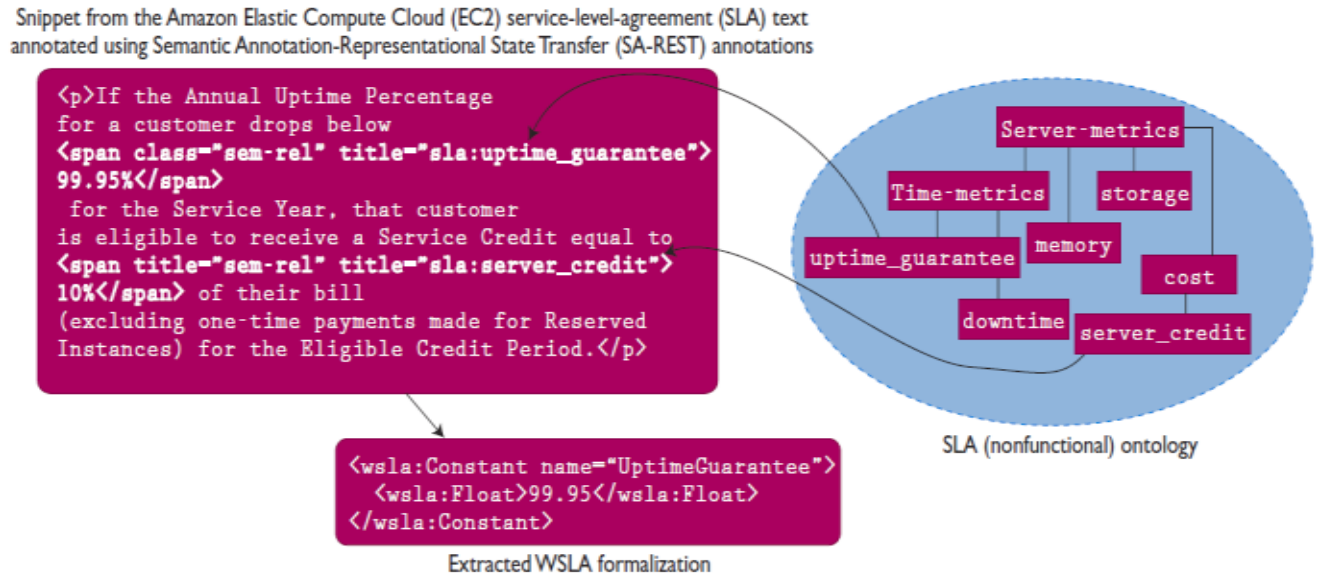

**Figure 9. Semantic annotations to embed machine-processable metadata to extract formal definitions [19].**

Similarly, as depicted in Figure 9, we can use semantic annotations to embed machine-processable metadata in order to extract formal definitions from structured text documents. In this example, a snippet from Amazon EC2 is being used to demonstrate how we can extract a formal definition of Web service-level agreement (WSLA) from EC2's human readable HTML document by simply annotating and linking the SLA components of the documents with corresponding SLA ontology.

Another notable work related to the issue of semantic models is the work presented by Mika and Tommerallo (2008) on Web semantics in the Clouds [20]. In that paper, the authors discussed a number of critical issues on deploying Cloud computing for the web of data. A system called 'Sandice' was developed which could potentially deal with large scale processing of structured web data. The Sandice system was developed by the Digital Enterprise Research Institute’s Data Intensive Infrastructures group (http://di2.deri.ie). The ultimate goal of the project was to develop scalable API that can be used to locate and utilize the massive amount of data available on the Web.

The example use cases for the Sandice API, as mentioned in the paper [20] include using keywords and semantic-pattern based queries to search for people, places, events, and connections based on semantically structured documents on the Web. These Web documents may include FOAF (friend-of-a-friend) RDF files, HTML pages with microformats and metadata, and XML Social Network Information (XFN). The Sandice system is designed to deal with three non-functional requirements: scalability, runtime performance, and the ability to cope with the frequent changes in standards on the Web of data [20]. Figure 10 depicts the internal architecture of Sandice system as extracted from extracted from Mika and Tommerallo (2008) [20]. As can be seen in Figure 10, the documents are first discovered and crawled from the Web as part of the indexing pipeline. The Sandice pipeline then extracts data from the Web documents and performs reasoning against the Sandice Ontology repository.

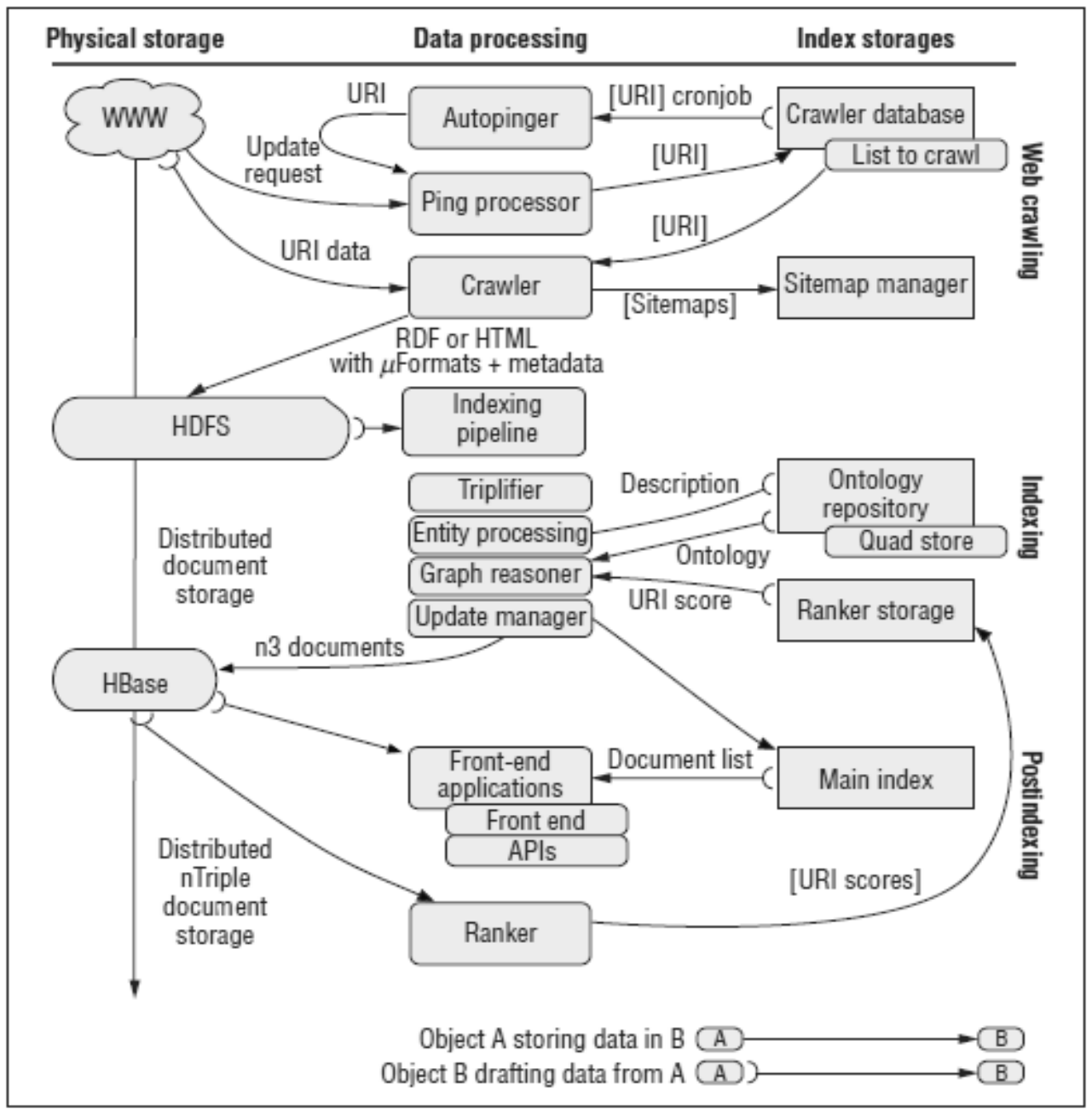

**Figure 10. The Sandice architecture as extracted from Mika and Tommerallo (2008) [20].**

The Sandice system then performs entity consolidation based on the ontological reasoning and creates the main index that can be used for the front-end applications and the APIs for the Clients. Hadoop Distributed File System (HDFS) was used as the distributed document storage for generated crawled data. The data from the HDFS are drafted to the Sandice indexing pipeline. The nTriple (similar to RDF) documents generated by the update manager are stored to the HBase system. The Sandice Ranker then use the data drafted from HBase to calculate appropriate ranking and passes the ranking data to Sandice's Ranker storage.

Note that the heart of the Sandice indexing pipeline is the ontological reasoning which has been applied to generate appropriate indexing for the RDF documents. Using the Sandice system, Mika and Tommerallo (2008) exploits the Cloud computing techniques to deal with the explosion of the Web of data [20]. As a concluding remark of their paper, the authors expressed their genuine interests in developing ontology-based Semantic web algorithms that can be transformed into well known Map/Reduce like frameworks. They urged the research community to explore the entire range of Semantic Web algorithms that could be transformed into Map/Reduce like solution space. The authors hoped that many of the scalability issues of the Semantic web can be solved in that way.

## 4.3 Cloud Computing Security

There exist a number of efforts that suggests the benefits of applying Ontologies to specify the security models for the Cloud computing services. One of the most cited efforts in this direction is the paper on Cyber security in Cloud Computing by Takahashi et al. (2010) [21]. In that paper, the authors proposed an ontological approach to achieve cyber security in the Cloud environment. The authors first developed an ontology for typical cyber-security operational information based on the actual operations in non-cloud environment as practiced in Japan, US, and South Korea. The authors then proposed to apply and extend their ontology to incorporate concepts and relationships that were found to be specific to Cloud computing cyber-security. Figure 11 depicts the proposed ontology that incorporates cybersecurity operational domains, entities, cybersecurity operational information along with their relationships.

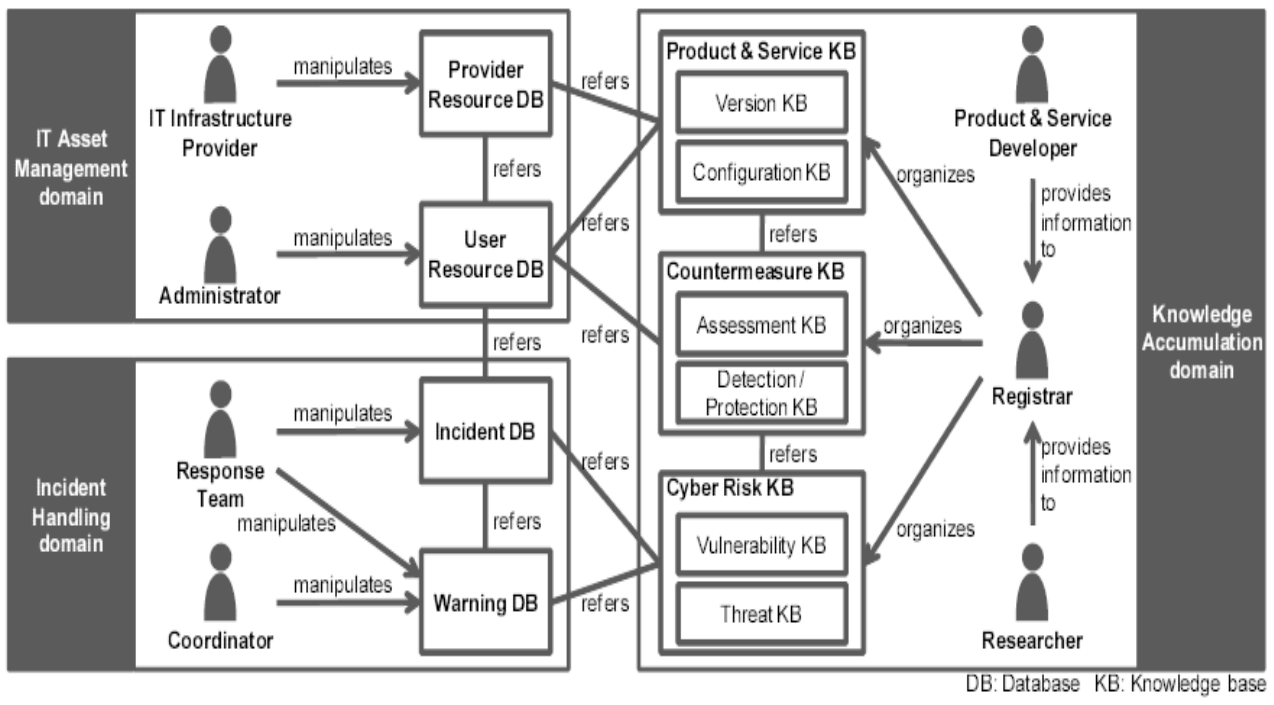

**Figure 11. The ontology of cybersecurity operational information proposed by Takahashi et al.(2010).**

The authors provided a clear set of requirements and detailed specifications on each of the operational domains, entities, and operational information of the proposed ontology. The proposed ontology is designed to be mapped with major cyber-security information standards that corresponds to the information categories suggested in Table 2. Based on their proposed ontology and corresponding cybersecurity information standards, the authors outlined the necessary information for cybersecurity in cloud computing. The authors also urged the need for standard description formats for such information for each type of the databases and knowledge bases in Table 2, and discussed their rationales. Note that a knowledge base represents the mapping between the data and their corresponding ontological concepts and entities.

Finally, based on the discussion of the paper [21], Takahashi et al. (2010) identified a set of necessary changes of cybersecurity information specific to the cloud computing environment. The authors rightfully argued that the three major factors for the changes are: 1. data-asset decoupling, 2. composition of multiple resources and 3. external resource usage. Note that the three factors considered by the authors are indeed the key distinctive characteristics of any Cloud computing environment (as contrary to traditional, non-cloud based computing).

**Table 2. Major Cybersecurity Operation Domains and Information Categories suggested by Takahashi et al. (2010).**

| Operation Domain | Information categories | |
|---|---|---|
| IT Asset Management | User Resource Database | |
| | Provider Resource Database | |
| Incident Handling | Incident Database | |
| | Warning Database | |
| Knowledge Accumulation | Cyber Risk Knowledge Base | Vulnerability Knowledge Base |
| | | Threat Knowledge Base |
| | Countermeasure Knowledge Base | Assessment Knowledge Base |
| | | Detection/Protection Knowledge Base |
| | Product Knowledge Base | Version Knowledge Base |
| | | Configuration Knowledge Base |

Another notable effort that employs ontology for Cloud computing security was presented in [22] by Bertram et al.(2010). In that paper, the authors introduce a new kind of Platform-as-a-Service (PaaS) architecture for the cloud which deploys semantic security risk management tools with dynamic web service policy frameworks. As discussed in the paper, the platform addresses the following key needs for the Cloud [22]: model security requirements, dynamically provision and configure security services, link operational security events to vulnerabilities and impact assessments. The authors evaluated their platform using a collaborative engineering design scenario deployed at a multi-tenant cloud. The platform is claimed to support security concerns throughout the whole lifecycle of a service-oriented Cloud based applications. Figure 12 depicts the on-demand Cloud security architecture as extracted from [22].

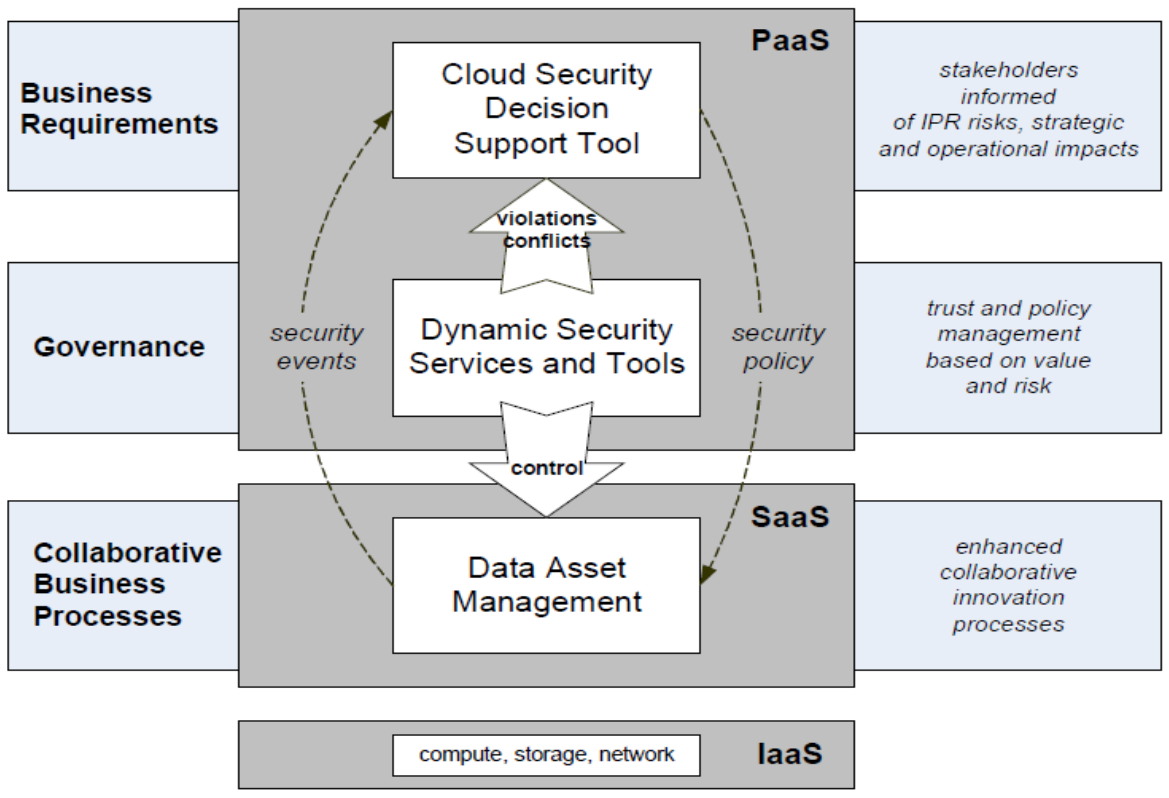

**Figure 12. On-demand Cloud security service architecture proposed by Bertram et al. (2010) [22].**

Note that the key component of the architecture in Figure 12 is the Security modeling and decision support tool. In order to achieve the decision support mechanism based on the security model, the authors employed the InfoSec ontology [23] to reason about the possible security risks within the underlying web service policies. Figure 13 depicts the key fragments of the InfoSec ontology that can be used to specify administration domains, communication links, and associated threats and vulnerabilities. Please refer to [23] for detailed descriptions for each of the classes depicted in Figure 13. This paper [22] by Bertram et al. (2010) demonstrated how ontologies can be utilized to model, analyze, plan, and monitor system security requirements within the cloud systems.

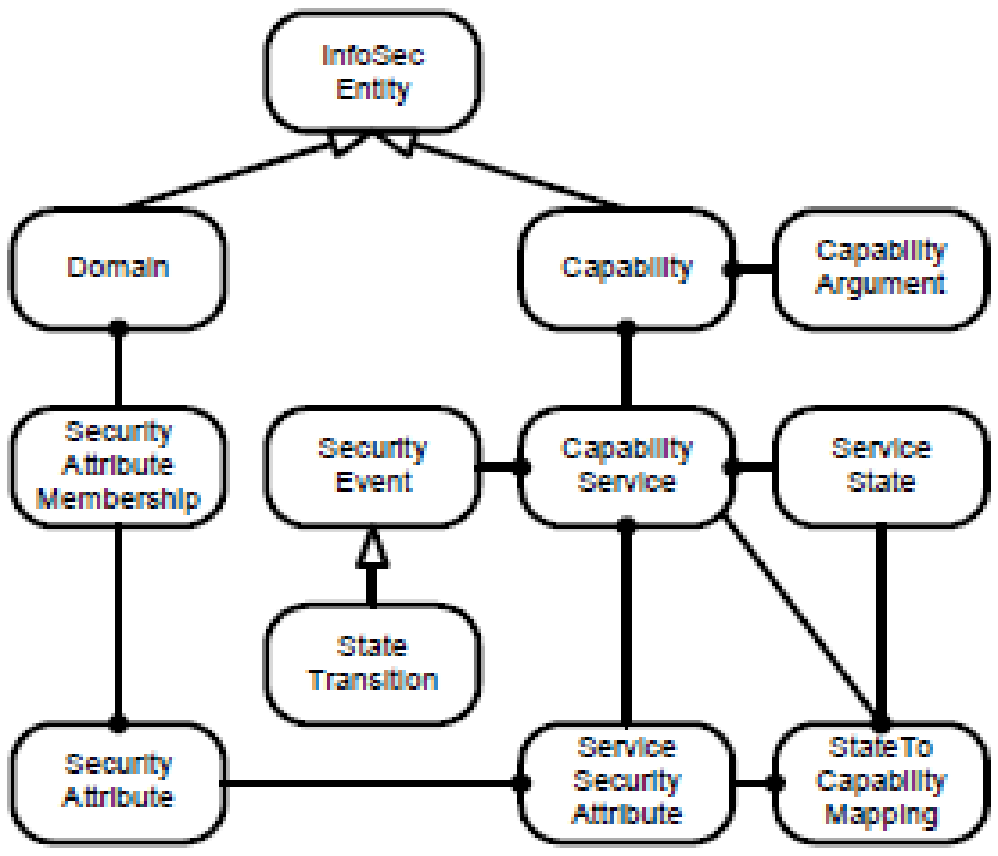

**Figure 13. Key classes of InfoSec ontology.**

Martinez et al. (2010) [24] presented a Cloud computing based malware detection system called uCLAVS (University of Caldas' AntiVirus Service) that exploits the strength of ontological reasoning. Their approach was based on the utilization of Malware Intrusion Ontology. The authors evaluated the effectiveness of uCLAVS using 1.2 million samples and ~25,000 malware instances.

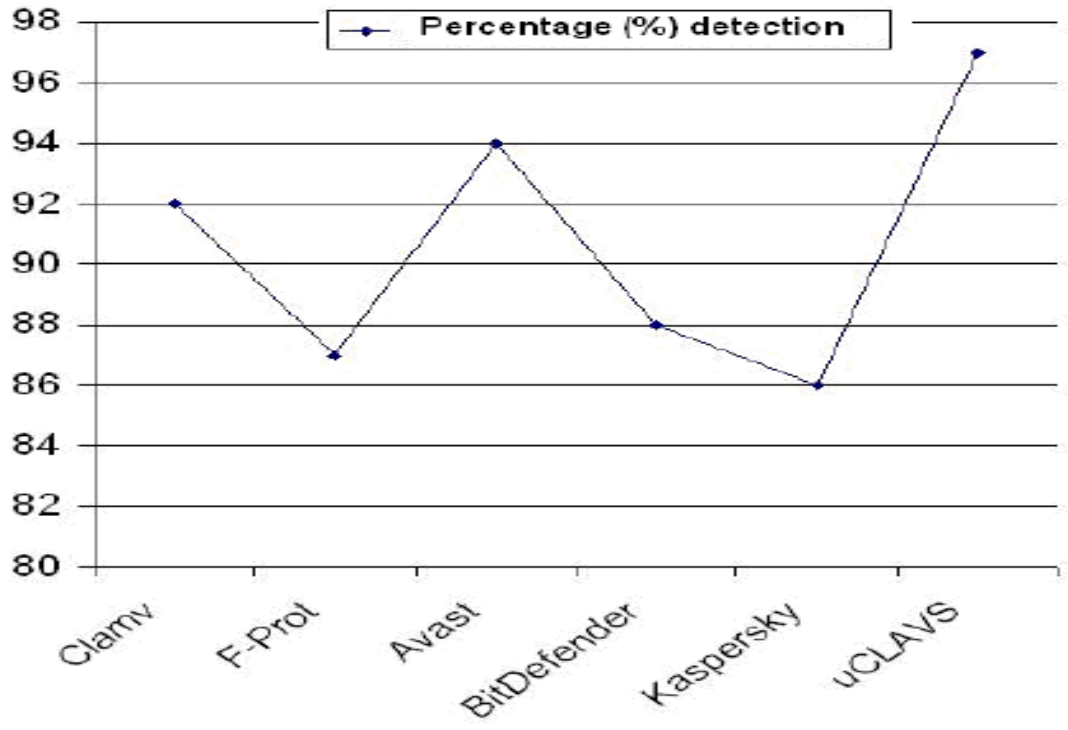

**Figure 14: uCLAVS Detection rate compared to other Anti-Malware Engines as presented by Martinez et al. (2010).**

As depicted in Figure 14 and Figure 15, the ontology-based uCLAVS approach significantly outperformed the most popular anti-malware engines.

| AntiVirus | 1 month | 1 week |
|---|---|---|
| Avast | 54,2% | 51,1% |
| AVG | 84,4% | 82,2% |
| BitDefender | 81,2% | 79,3% |
| ClamAV | 56,7% | 54,2% |
| F-Prot | 53,4% | 51,2% |
| Kaspersky | 89,1% | 86,8% |
| ***uCLAVS*** | ***97,07%*** | ***93,4%*** |

**Figure 15. Detection Rate with the age of Malwares using different Antivirus as presented by Martinez et al. (2010).**

## 4.4 Cloud Resource Management and Cloud Service Discovery

One of the concrete examples of applying ontologies for automated resource management for the Cloud has been presented by Ma and Jeng et al. (2011) in [25]. The paper proposed an ontology-based job allocation algorithm for a resource management system in cloud environment with very promising results.

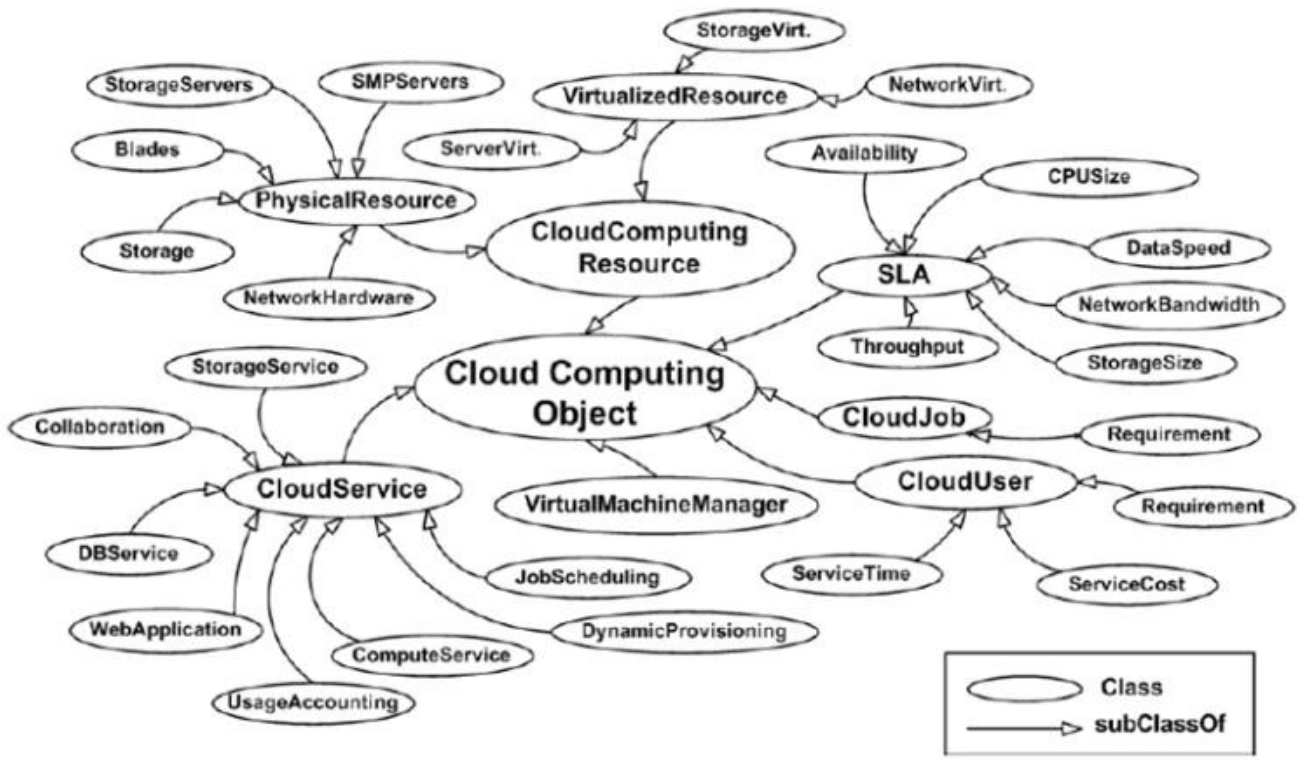

**Figure 16. Upper level skeleton of the Cloud Resource Ontology developed by Y. Ma et al. (2011) [25]**

As part of their system, the authors developed an OWL ontology (Figure 16) to represent different cloud resource information and Service Level Agreements (SLAs) along with their semantic meaning. The approach guarantees Quality of Service (QoS) by dynamically allocating cloud resources for the users based on the agreed SLAs. Figure 17 provides the high level architecture of the ontology-based resource allocation system as extracted from [25]. As depicted in the figure, in the heart of the system is the Cloud ontology and OWL/RDF rule engine which takes as inputs the information from Could users and Cloud service providers in terms of their negotiated SLA. The system then executes a set of rules to instantiate all the relevant classes of the ontology (e.g., Cloud User, OS type, CPU Size, Data Speed, and Storage Size etc.) and provides a set of logically inferred results for appropriate cloud resource allocation.

The logical rules used for the system are written in a language called SWRL (Semantic Web Rule Language) which can be applied to any OWL/RDF based ontologies. Six examples of such rules as used by the system proposed in [25] are provided in Table 3. The SWRL rules can take advantage of OWL reasoners and instantiate different classes (with individual data) that follow the specified rules written in First Order Logic Language (FOL) like syntax. For example, rule 3 in table 3 describes the rule to instantiate all the candidate resources. The rule essentially reads as: for any ?x of type Resource, if ?x is an Available Resource *and* ?x has the Service Type ?a *and* ?a satisfies the quality of required service type, *then* instantiate ?x as a Candidate Resource in the ontology. The other rules described in Table 3 can be read in a similar fashion.

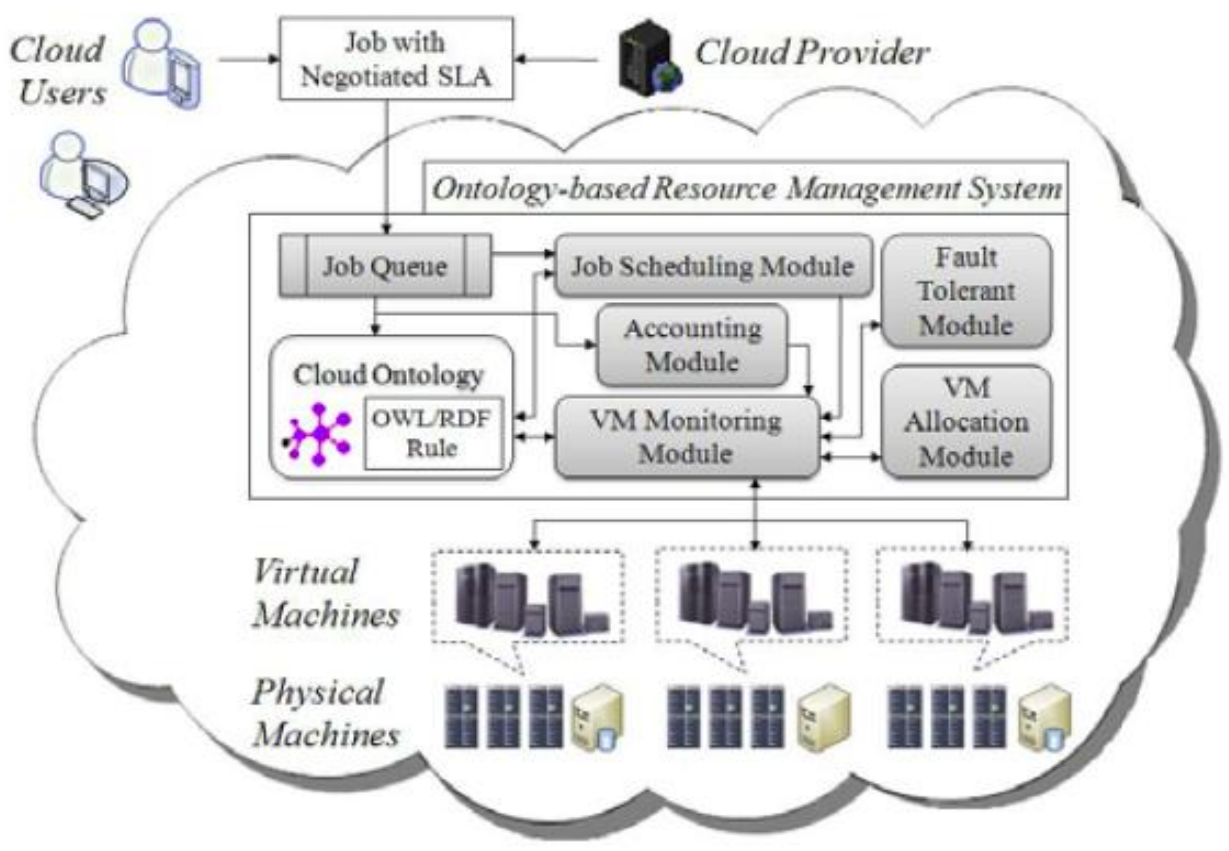

**Figure 17. Architecture of the ontology-based resource management system proposed by Y. Ma et al. (2011) [25].**

The authors compared their proposed Ontology-based Resource Management Algorithm (ORMA) with two existing resource management algorithms such as Profit-Optimization-based Resource Management Algorithm (PORMA) and the Response time-Optimization-based Resource Management Algorithm (RORMA). A simulation model was developed by applying the discrete event system specification (DEVS) formalism [26] for their ontology-based resource management system. The evaluated results indicate that overall, the proposed ontology-based approach clearly outperformed the other two existing approaches in terms of average throughput and resource utilization.

**Table 3. The SWRL rules for the Cloud Resource Ontology.**

| No | Domain rules |
|---|---|
| 1 | VirtualizedResource(?x)∧registeredToVMM(?x, ?y)∧VirtualMachineManager(?y) ∧groupName(?z) → includedInVMM(?x, ?z) |
| 2 | VirtualizedResource (?x) ∧expectTime(?x, ?a) ∧[?a<=qDeadline]∧expectCost(?x, ?b)∧[?b<=qBudget] → AvailableResource(?x) |
| 3 | AvailableResource(?x) ∧hasServiceType (?x, ?a) ∧[?a=qServiceType] → CandidateResource(?x) |
| 4 | VirtualizedResource (?x)∧osType(?x, ?a)∧[?a=windowXP∨windowNT]∧cpuSpeed(?x, ?b)∧[?b>=1.80]∧ramSize(?x, ?c)∧[?c>=1024]∧bandwidth(?x, ?d)∧[?d>=100]→ hasServiceType(?x, ComputeService) |
| 5 | VirtualizedResource (?x)∧osType(?x, ?a)∧[?a=Window2000Server∨Window2003Server]∧responseTime(?x, ?b)∧[?b>=3.0]∧cost(?x, ?c)∧[?c>=200]∧storageSize(?x, ?d)∧[?d>=100]→ hasServiceType(?x, WebApplication) |
| 6 | VirtualizedResource (?x)∧osType(?x, ?a)∧[?a=Linux]∧cpuSpeed(?x, ?b)∧[?b>=2.13]∧cost(?x, ?c)∧[?c>=100]∧storageSize(?x, ?d)∧[?d>=1000]→ hasServiceType(?x, StorageService) |
| … | … |

The results are provided in Figure 18 as extracted from [25]. The dark dotted curves represent the performance of the ORMA approach. As the result suggests, on average ORMA provided the throughput of little over 5.0 jobs, which is higher than that of the other two algorithms. Although ORMA's average throughput is comparable with RORMA, ORMA did process more jobs than other algorithms. The improvement in average resource utilization through ORMA is quite impressive compared to the other two approaches. On average, ORMA provided 32.3% and 13.7% higher resource utilization than PORMA and RORMA respectively. This result indicates that ORMA allocates jobs more uniformly than the existing resource management algorithms [25].

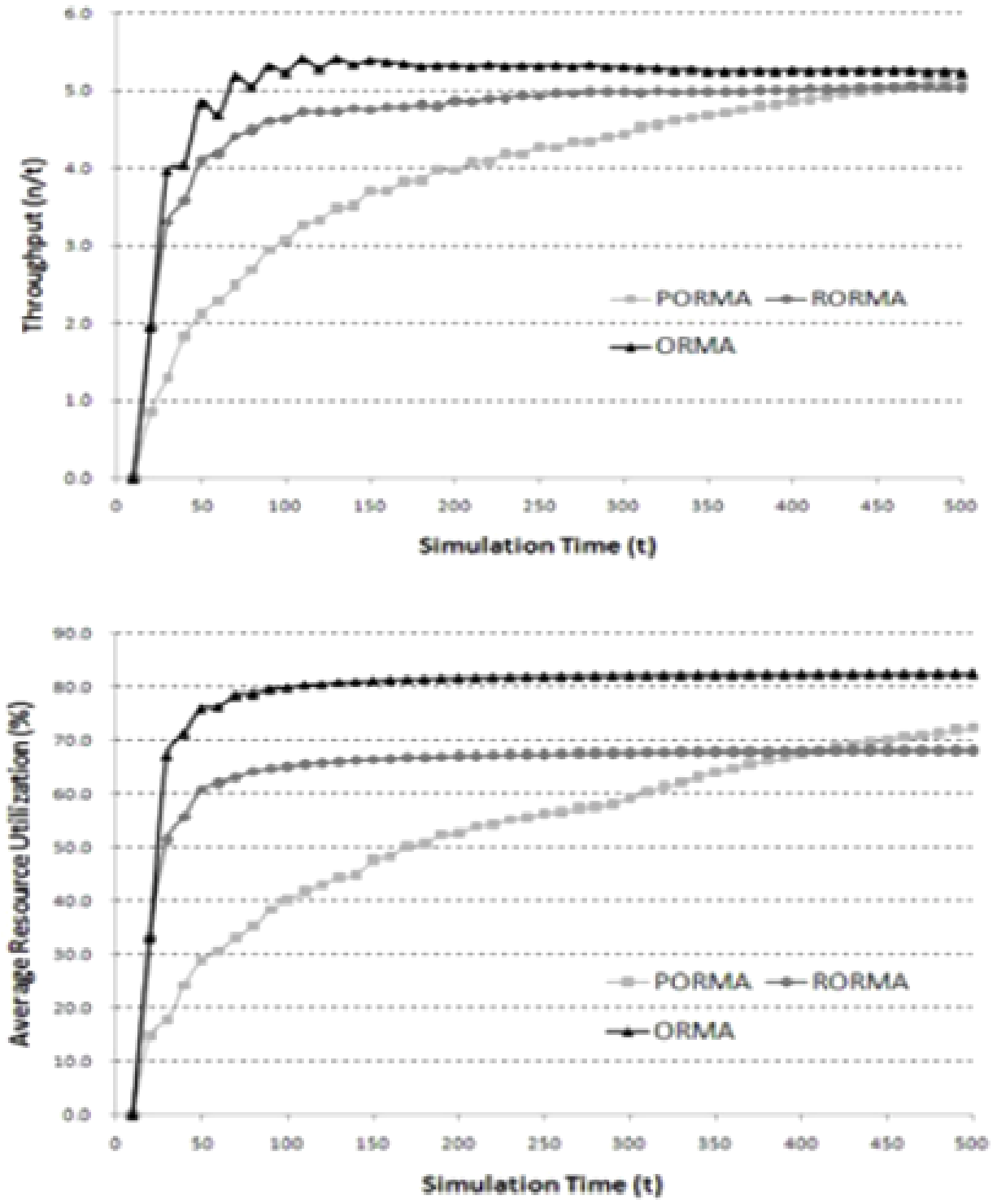

**Figure 18. Performance measures for ORMA compared to existing resource management algorithms [25].**

Han and Sim (2010) present an intriguing work on an ontology-enhanced service discovery system for the cloud environment in [27]. The paper presents a system called Cloud Service Discovery System (CSDS) which can assist the clients to find appropriate Cloud services over the internet. As a critical component of the CSDS project, the authors developed a Cloud Ontology (CO). The ontology is used to assist the CSDS system to determine similarities among different available Cloud services. One of the significant aspects of the project is that it has been claimed as the first attempt of developing an agent-based knowledge discovery system which utilizes an ontology to retrieve information about Cloud services.

As depicted in Figure 19, the core component of the overall CSDS system is the Cloud Service Reasoning Agent (CSRA) in the middle. By interacting with the Cloud Ontology, the CSRA enables the main functionalities of the CSDS. The reasoning engine within CRSA first passes the Cloud terms (as retrieved from the web portals of the cloud providers) to reason against the Cloud ontology about the relational nature of the available Cloud services. Based on the similarity score of each of the service terms within the ontology, the CRSA reasoner returns the rating of the search results. The Cloud ontology developed for the CSDS project contains the concepts of different Cloud services.

The ontology enables the CSRA to determine the relations of Cloud services using three service reasoning methods as described in [27]: 1) Similarity reasoning, 2) Equivalent reasoning, and 3) Numerical reasoning.

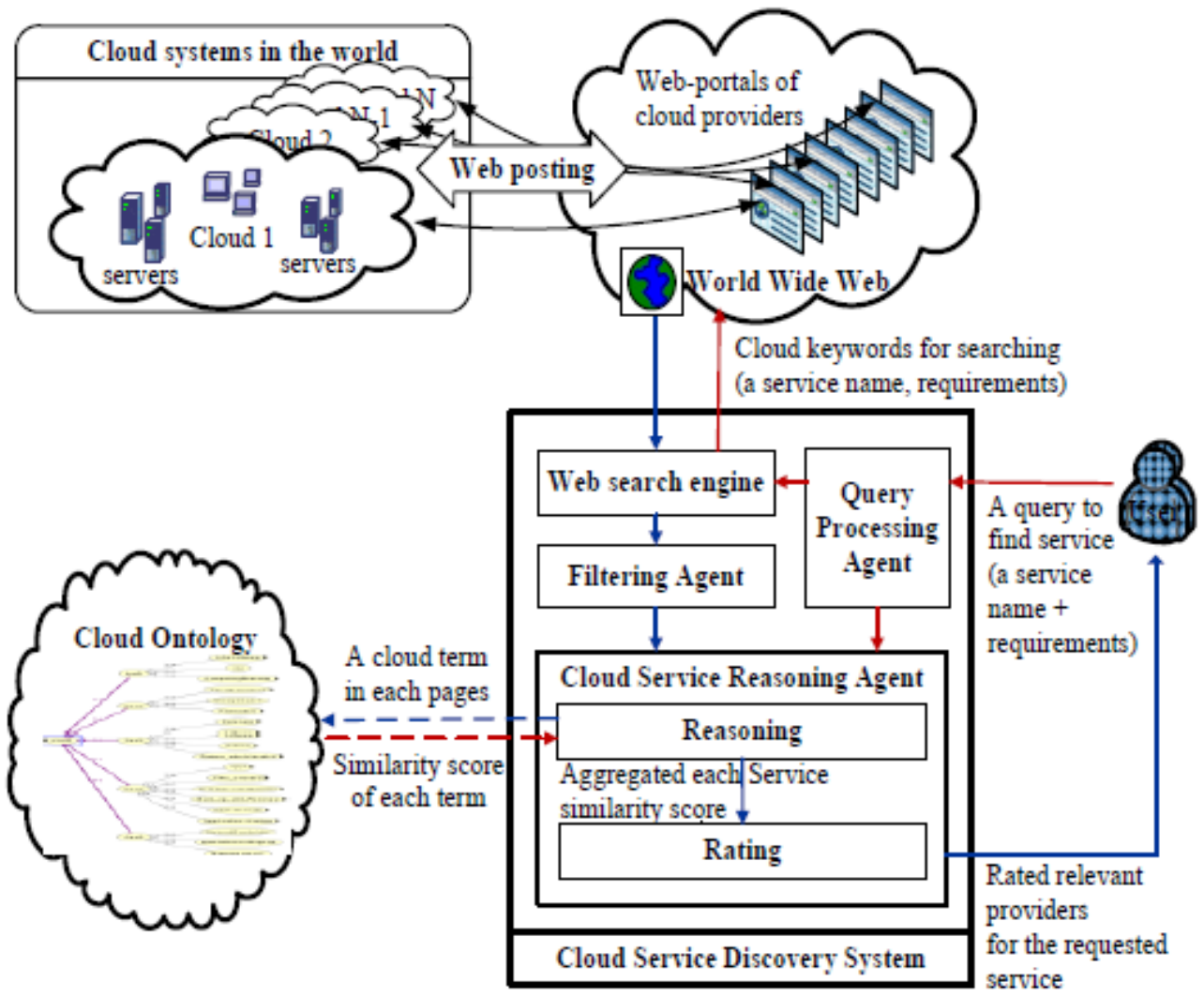

**Figure 19. Ontology based Cloud Service Discovery System (CSDS) proposed by Han and Sim (2010) [27].**

In their paper, the authors only provided the algorithm for the Service similarity reasoning. The other two reasoning algorithms for the CSRA were left for a future paper. The authors provided the detailed results of their empirical study on CSDS using the experimental settings in Table 4.

**Table 4. Experimental settings for Cloud Service Discovery System (CSDS) simulations by Han and Sim (2010) [27]**

| Experiment variables | Value (range) |
|---|---|
| The number of providers | 40, 70, 100, 130, 160, 190 |
| The number of Cloud services provided by each provider | 25~35 (web-pages) |
| The number of Cloud service web-pages in the virtual-www | 1200, 2100, 3000, 3900, 4800, 5700 (web-pages) |
| The number of web-pages in the virtual-www (not for Cloud service) | 10,000 web-pages |
| Total number of web-pages in the virtual-www | 11200, 12100, 13000, 13900, 14800, 15700 |
| The number of Cloud services | Around 100 service names |
| CPU clock | 0.1~6.0 GHz |
| RAM size | 0.256~36.0 GB |
| HDD size | 0.1~1000 GB |
| Network Bandwidth | 0.1~10 Gbps |
| Network Latency | 1~5000 ms |

The results indicated that using the Cloud ontology, the CSDS was more successful in finding Cloud services that were closer to the clients' requirements. The performance measures used by the authors are based on the results of *Service Utility* and *Success Rate* as depicted in Figure 20 and Figure 21 respectively.

As evident in Figure 20, in terms of service utility, CSDS *with* the Cloud ontology provided better performance than *without* the CSDS. As pointed out by the authors, the reason behind this performance improvement with the Cloud ontology is due to the filtering and ontological reasoning functionalities of CSDS which would return web-pages of the Cloud services with higher ratings that are more likely to be closer to clients' requirements.

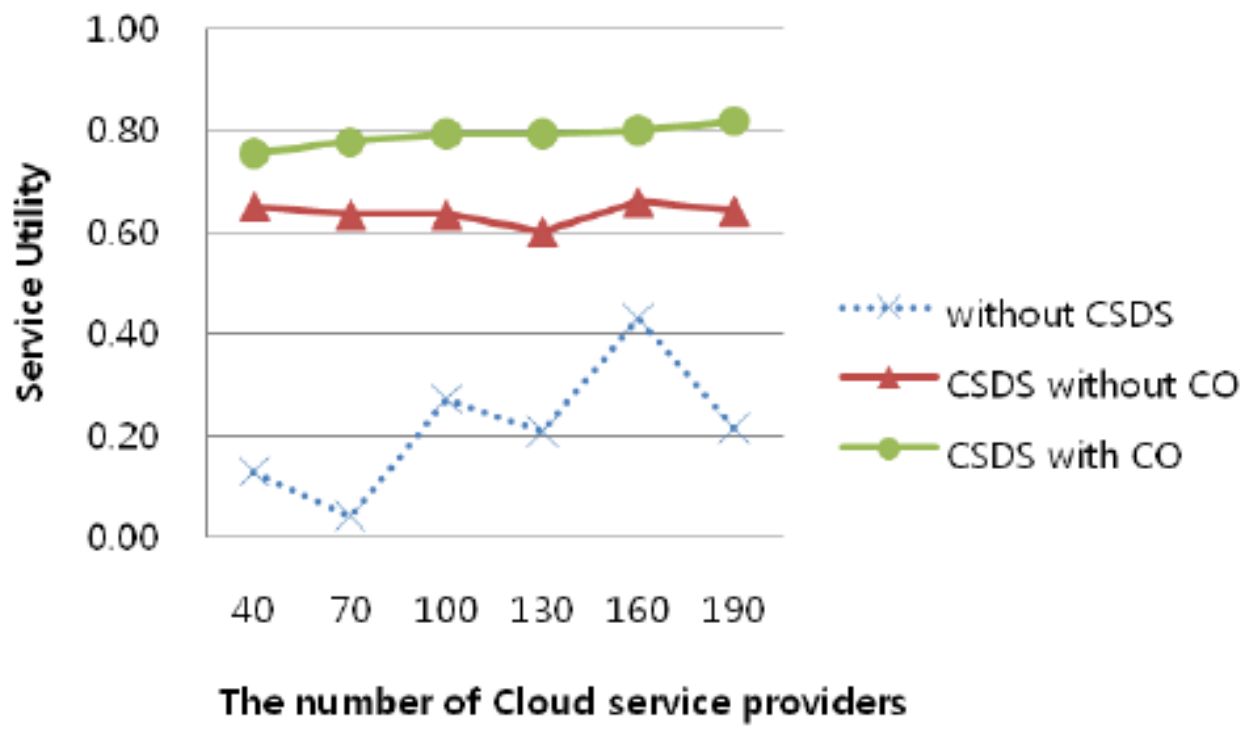

**Figure 20. Service Utility evaluation without CSDS, and CSDS with and without Cloud Ontology (CO) as presented by Han and Sim (2010) [27].**

As for the success rate (calculated as the ration between the *number of success* and the *number of attempts*), the assumption asserted was that a service discovery would fail if the service utility was less than 0.5. The result in Figure 20 shows that the retrieved Web pages using CSDS with the Cloud ontology had service utilities which were well over the margin of 0.5. As depicted in Figure 21, the resulting plots clearly indicate that the users were more successful in discovering their desired Cloud services using the CSDS *with* the Cloud ontology compared to the CSDS *without* the Cloud ontology.

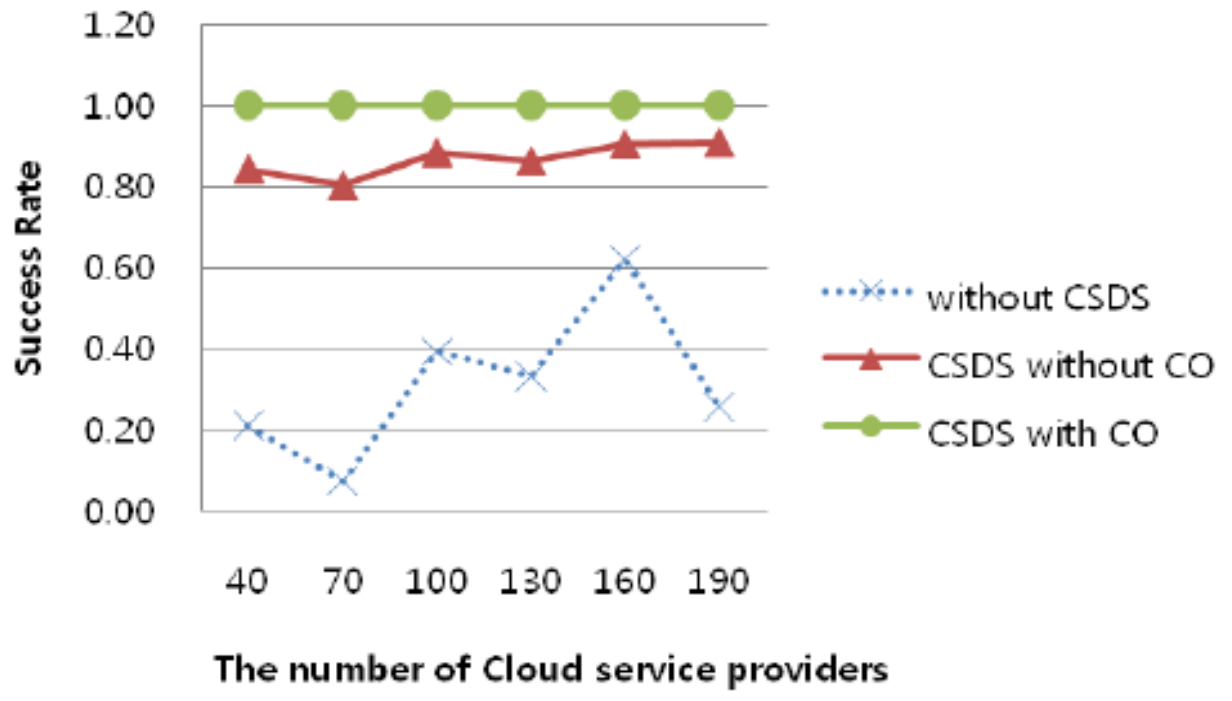

**Figure 21. Success Rate evaluation without CSDS, and CSDS with and without Cloud Ontology (CO) as presented by Han and Sim (2010) [27].**

Another notable effort associated with Cloud resource selection service that required extensive use of ontologies along with their reasoning capabilities was presented by Yoo et al. (2009) in [28]. This paper proposed an ontology-based resource virtualization mechanism for the Cloud that could be used specifically for the Science cloud. As part of the project, the authors introduced the idea of Virtual Ontology (VOn) that can be dynamically configured based on the clients' requirements. The conceptual entities from the VOn are then mapped to different Cloud based resources. The service proposed by the authors employs a Map/Reduce model to efficiently merge a number of relevant ontologies and calculate the rapid ranking of the available resources. As depicted in Figure 22, the architecture of the ontology-based Resource Selection System (OReSS) has four major layers [28]: Physical Machines (PM) layer, Cloud Resource Virtualization (CRV) layer, OReSS layer, and End User (EU) layer. Refer to [28] for the detailed description of each of the layers.

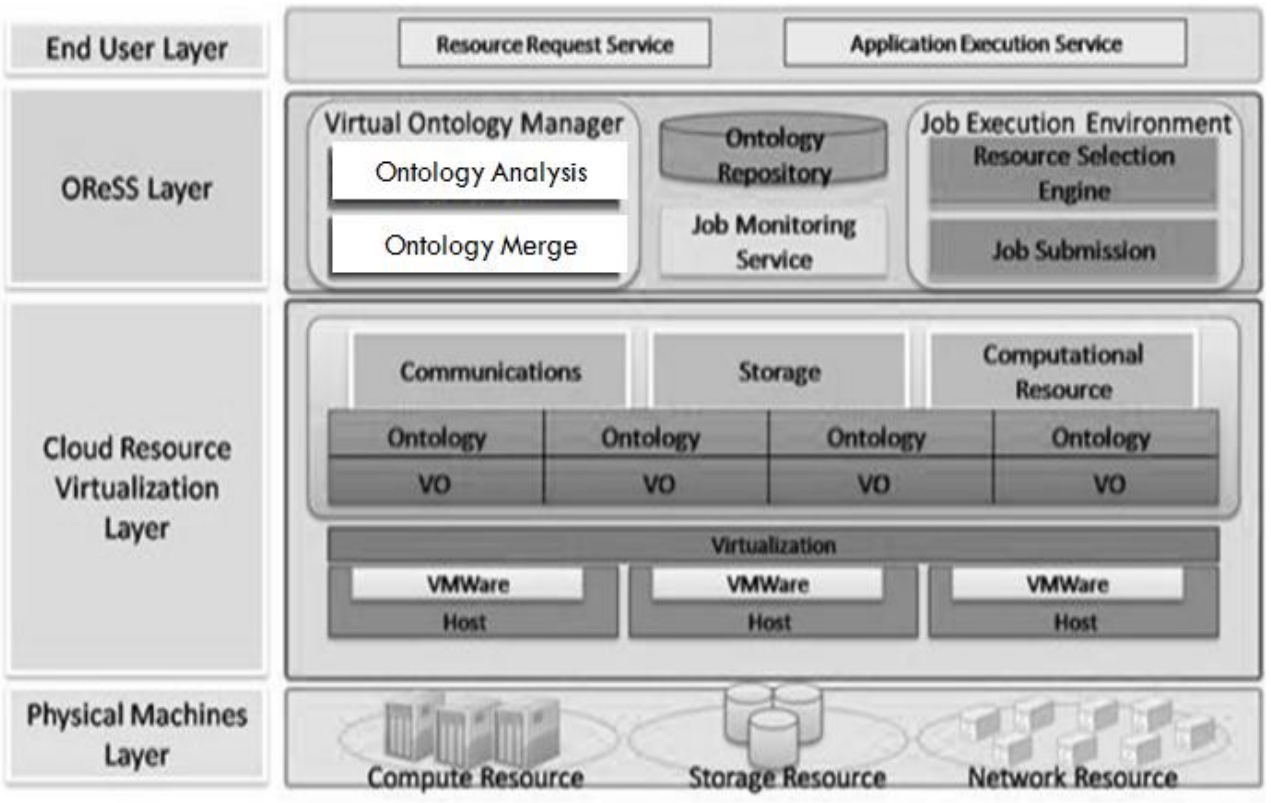

**Figure 22. The architecture of the Ontology-based Resource Selection Service (OreSS) proposed by Yoo et al. (2009) [28].**

The core strength of the resource selection mechanism in OReSS is the Ontology Analysis engine that calculates similarity to clients' requirements. Based on the similarity measures, the system then derives the ranking of the ontologies from the CRV layer of the OreSS. Once the rakings of the ontologies are computed, the Ontology Merge engine generates a new VOn by combining the candidate set of ontologies based on their degree of similarity received from the clients. The merged ontology would then be considered as the closest resemblance of the specification of the required Cloud resources. Once the resources are scheduled for executing jobs, the proposed system considers the following three factors [28]: generated VOn from Ontology merging engine, status of the stored resources within the VOn's resource pool, and the resource requirements from a Client. Finally, if the latter three constraints are met satisfactorily, the scheduled jobs can be run on the selected resources. As we can observe the mechanism used by OreSS clearly depends on the use of ontologies and their reasoning mechanisms.

## 5. Discussion

The overall idea of Cloud computing has been evolving ever since the inception of the term itself. There exist no unified way to identify and classify all the aspects of Cloud computing phenomena that can support common vocabulary and common understanding among the cloud computing researchers, as well as the industries that provide Cloud based services. One of the reasons for this is that the idea of "Cloud computing" significantly overlaps a number preceding ideas from different fields in computer science, specially from distributed systems, database management systems, parallel computing architectures, web based computing, and so forth. The idea of having an ontology that could represent the overall domain of Cloud computing in a uniform way, therefore, is quite intriguing. However, as observed in Section 4, we can see a number of research efforts that offers ontology-based solutions to confront some of the other, more critical issues in Cloud computing. In this section, we discuss the efforts surveyed in this paper in a qualitative manner. Figure 23 in the next page provides the summary of the surveyed efforts in terms of their achievements and possible limitations.

### 5.1 Cloud Systems Interoperability

The potential benefits of using ontologies to achieve semantic interoperability among the heterogeneous Cloud systems have been well established in [9], [18], and [5]. The layered architecture proposed by Youseff et al. (2008) [9] has formed the basis of a unified ontology for the overall phenomena of Cloud computing. The hope of the paper was to promote much desired interoperability among the Cloud systems. The paper describes the fundamental properties of Cloud computing in a systematic manner. With over 400 citation records, this work has been very influential among the Cloud computing and ontology researchers. Although the high-level architecture presented by Youseff et al. (2008) forms the basis for developing a comprehensive ontology, the work is quite limited in terms of addressing the formal, logical aspects of ontologies. However, it should be noted that the work is important for anyone who wants to gain a comprehensive understanding on the essential components of Cloud computing.

When it comes to actual formal specification of Cloud computing concepts, the efforts by the EU based mOSAIC project [18] is quite impressive. The comprehensive mOSAIC ontology does have the potential to effectively represent the overall domain of Cloud computing. The impact of the mOSAIC project can be observed through the cluster maps of their main research site (http://www.mosaic-cloud.eu/). The mOSAIC project page has been visited over 14,000 times since the inception of the project in August 2010. This is quite impressive for a scientific project which could gain attention from the research communities all over the globe (see the analytics at http://goo.gl/DtCdI). Although the overall effort of the mOSAIC project is quite impressive, its dependence on Semantic web technologies may cause severe performance issues for ontological reasoning services for their query processing engine. This is a severe limitation of the mOSAIC platform which was simply overlooked in [18]. So far, to our best knowledge, none of the semantic web tools provide scalable solution to the issue of ontological reasoning. Unless the reasoning services can be cast into Cloud-based Map/Reduce like solutions as suggested in [20], the true benefit of semantic web technologies for the Cloud environment may never be realized in actuality.

The effort suggested in UCI project [5] can also suffer from the same limitation of scalability as in mOSAIC platform. Also, it should be noted that both UCI and mOSAIC project shares almost identical goals of having a universal platform and API for multi-cloud systems. This kind of duplication of efforts must be avoided, and it would be more beneficial for the community if UCI and mOSAIC project could unify their ontologies into a single one.

### 5.2 Semantic Models and the Cloud

As observed in Section 4.2, Sheth et al. identifies three major aspects of Cloud computing where semantic models could be helpful. The aspect of semantic interoperability between Cloud systems may suffer the analogous issues discussed in 5.1. The proposed approach of semantic annotation of DSL script or structured documents can indeed be very useful to derive the higher-level abstractions of an application. However, the author did not discuss any of the practical issues to automate the process of annotation and mappings with ontologies. The proposed approach may require a vast amount of manual labor in order to annotate the specific fragments of different source code document.

The Sandice system discussed by Mika and Tommerallo (2008) [20] to process the web of data provides a number of practical issues on scalability that are usually overlooked by the Semantic web community. The approach is quite practical as it takes advantage of Hadoop Map/Reduce solutions to process the vast amount of data from RDF triples. Although the approach looked very practical, the authors did not provide any performance measure that could demonstrate the strength of their approach. Also, it would be extremely difficult to implementing a framework like Sandice in a non-industrial setting.

### 5.3 Cloud Computing Security

The benefits of using ontologies for the security aspects in Cloud environment have been well recognized by the efforts observed in Section 4.3. However, the approaches discussed in that section should require further, practical evaluation. The proposed ontology to represent the concepts and entities of cybersecurity for the Cloud environment by Takahashi et al. (2010) looked very promising. However, when it comes to using the ontology to see the actual benefit, we must map different concepts from different security protocols that are used in different Cloud systems. The process of mapping the security protocols from different Cloud systems would certainly require a vast amount of manual labor. The authors did not seem to have much concern on this practical issue.

The ontology-based decision support tool used by Bertram et al. (2010) for on-demand Cloud security service looked promising as well. However, the core of their approach, the *dynamic* decision support system provokes some skepticism. The decision support system proposed by the author directly relies on ontological reasoning. This kind of on-the-fly reasoning may cause severe performance issues.

The performance evaluation presented for the Cloud-based malware detection system developed by Martinez et al. (2010) [24] is quite impressive. However, the experimental setup seemed too complex to re-implement if we wanted to re-evaluate or verify the results.

**Cloud Systems Interoperability**

- Significant awareness to recognize the importance of ontologies.
- Yuseff et al. (2008) [9]
  - Proposed a high level architecture for an unified ontology for the overall paradigm of Cloud computing.
  - *Limitation:* Formal aspects of ontologies were not considered.
- The EU based mOSAIC project [18]
  - Developed a common ontology compliment with existing standards, platform and services
  - Provides an ontology-based open source API and platform for the prominent multi-cloud environments
  - *Limitation:* Poor scalability of the reasoning services may hinder the overall goal of the project.
- The Unified Cloud Interface (UCI) project [5]
  - Aims at employing Semantic web technologies to resolve Cloud interoperability issues.
  - Aims at providing a *single* interface (API) to access and retrieve all multi-cloud resources in a unified way
  - *Limitation:* Too ambitious at this moment. Semantic web tools and ontology reasoners are still not ready to deal with large-scale systems

**Semantic Models and the Cloud**

- Identified the aspects to Cloud where semantic models can help.
- Sheth et al. (2010) [19]
  - Proposed a technique of annotating DSL scripts of Cloud APIs with high-level ontological semantic models
  - Semantic annotation on structured web documents to extract the formal ontological definitions
  - *Limitation*: May require vast amount of manual interventions for annotating the ontology mappings
- Mika and Tommerallo (2008) [20]
  - Developed an ontology-based framework called 'Sandice' to deal with large scale processing of structured web data using Map/Reduce paradigm on Semantic web reasoning
  - Developed a scalable API to locate and utilize the massive amount of data available on the Web
  - *Limitation*: Performance is not evaluated. Geared towards industrial applications.

**Cloud Computing Security**

- Benefits of ontologies have been recognized but requires more evaluation.
- Takahashi et al. (2010) [21]
  - Proposed an ontology for cyber-security operational information applicable to the Cloud environment.
  - Proposed ontology is mappable with major cyber-security information standards.
  - *Limitation*: May require tedious manual interventions to map the security protocols with different cloud systems.
- Bertram et al.(2010) [22]
  - Employs an ontology-based security models and decision support tool for an On-demand Cloud security service architecture
  - *Limitation*: Heavily depends on dynamic decision support. On the fly reasoning in large-scale may cause severe performance issues.
- Martinez et al. (2010) [24]
  - Developed an effective malware detection system (uCLAVS) that exploits reasoning on Malware Intrusion Ontology.
  - *Limitation*: Complex experimental setting. Would be hard to replicate the result.

**Cloud Resource Management and Service Discovery**

- The most promising field in terms of utilizing ontologies.
- Ma and Jeng et al. (2011) [25]
  - Developed an ontology-based job allocation algorithm for a Cloud resource management system.
  - Promising results that outperform existing approaches.
  - *Limitation*: Reasoning on the fly on the SWRL rules will have significant performance issue.
- Han and Sim (2010) [27]
  - Developed an ontology-enhanced service discovery system for the Cloud users.
  - Deployed an ontology-based cloud service reasoning agent with promising results.
  - *Limitation*: On the fly reasoning would hinder the performance.
- Yoo et al. (2009) [28]
  - Developed an ontology-based resource virtualization mechanism for the Science Cloud.
  - Deployed an ontology analysis engine to calculate the matching similarities based on clients' requirements.
  - *Limitation*: heavily depends on dynamic re-configuration and merging of virtual ontologies. This may have severe performance issue in large scale.

**Figure 23. Summary of the surveyed efforts.**

### 5.4 Cloud Resource Management and Service Discovery

In terms of applying ontologies and their reasoning capabilities, the aspects of Cloud resource management and cloud service discovery are probably the most intriguing ones. As observed in Section 4.4, the ontology-based approaches targeting these aspects of Cloud computing produced promising very performance results.

As depicted in Figure 18, the ontology-based resource allocation algorithm developed by Ma et al. (2011) [25], significantly outperformed the existing approaches. However, it should be noted that the evaluation was performed on an experimental settings where the reasoning results were pre-computed. If we want to run the system on the fly, the ontological reasoning on the SWRL rules used by the system may cause significant performance penalty. The ontology-based approach developed by Han and Sim (2010) [27] for Cloud service discovery system may also suffer from the analogous issue of compromised performance for dynamic service selection operations.

Finally, as described in Section 4.4, the resource virtualization system proposed by Yoo et al. (2009) [28] heavily depends on rapid re-configuration and merging of virtual ontologies (VOn). The process of dynamic re-configuration and merging of the virtual ontologies may entail the analogous issues of reasoning overheads as observed in the latter two systems.

## 6. Concluding Remarks

As the title of this paper suggests, this project was expected to result in providing the current state-of-the-art understandings of ontologies and their applications within the phenomena of Cloud computing. As the results of this survey suggests, ontologies along with their reasoning capabilities can have some significant impacts on improving or simplifying a number of critical aspects in Cloud Computing practices.

We have observed a number of specific aspects of Cloud computing where ontology-based solutions were applied, as presented in different literatures. The scope of this project was targeted towards the following four specific aspects of Cloud computing: Cloud Systems Interoperability, Semantic Models and the Cloud, Cloud Computing Security, and Cloud Resource Management and Service Discovery. It should be noted that the emergence of new paradigms like Linked Data [14], BigData [13], and Semantic Web technologies are also playing critical roles in applying ontologies for the Cloud computing paradigm. However, these relatively new aspects were not covered in this survey, and are left for future work.

Although this paper is written as a Cloud Computing course project, we expect to publish our findings in a refereed journal or a conference proceeding in near future. It is hoped that publishing this survey can potentially promote new research areas and possibilities for both ontology and Cloud computing communities.